\documentclass[12pt]{article}
\usepackage[margin=22mm]{geometry}
\usepackage{graphicx}
\usepackage[version=3]{mhchem} 
\newcommand{\equn}[1]{\begin{equation}\label{#1}}
\newcommand{\eqan}[1]{\begin{eqnarray}\label{#1}}
\newcommand{\eqa}{\begin{eqnarray}}
\newcommand{\equ}{\begin{equation}}
\newcommand{\nuqe}{\end{equation}}
\newcommand{\uqe}{\end{equation}}
\newcommand{\naqe}{\end{eqnarray}}
\newcommand{\aqe}{\end{eqnarray}}
\newcommand{\nonu}{\nonumber}

\newcommand{\la}{\langle}
\newcommand{\ra}{\rangle}
\newcommand{\e}{{\rm e}}

\usepackage{color}

\makeatletter
  \newcommand\reaction@[1]{\begin{equation}\ce{#1}\end{equation}}
  \newcommand\reaction@nonumber[1]%
    {\begin{equation*}\ce{#1}\end{equation*}}
  \newcommand\reaction{\@ifstar{\reaction@nonumber}{\reaction@}}
\makeatother

\begin{document}

\title{Lecture notes on stochastic models in systems biology}
\author{Peter S.\ Swain
\\ 
{\normalsize {\tt peter.swain@ed.ac.uk}}\\{\normalsize Biological Sciences, University of Edinburgh}}
\date{}
\maketitle

\begin{abstract}
  These notes provide a short, focused introduction to modelling
  stochastic gene expression, including a derivation of the master
  equation, the recovery of deterministic dynamics, birth-and-death
  processes, and Langevin theory. The notes were last updated around
  2010 and written for lectures given at summer schools held at McGill
  University's Centre for Non-linear Dynamics in 2004, 2006, and 2008.
\end{abstract}

\section*{Introduction}
A system evolves stochastically if its dynamics is partly generated by
a force of random strength or by a force at random times or by
both. For stochastic systems, it is not possible to exactly determine
the state of the system at later times given its state at the current
time. Instead, to describe a stochastic system, we use the probability
that the system is in a certain state and can predict how this
probability changes with time. Calculating this probability is often
difficult, and we usually focus on finding the moments of the
probability distribution, such as the mean and variance, which are
commonly measured experimentally. 

Any chemical reaction is stochastic. Reactants come together by
diffusion, their motion driven by collisions with other
molecules. Once together, these same collisions alter the internal
energies of the reactants, and so their propensity to react. Both
effects cause individual reaction events to occur randomly.

Is stochasticity important in biology? Intuitively, stochasticity is
only significant when typical numbers of molecules are low. Then
individual reactions, which at most change the numbers of molecules by
one or two, matter. Low numbers are frequent {\it in vivo}: gene copy
number is typically one or two, and transcription factors often number
in the tens, at least in bacteria. There are now many reviews on
biochemical
stochasticity\cite{Kaern:2005gr,Shahrezaei:2008ez,Raj:2008ip,Eldar:2010kk}.

Unambiguously measuring stochastic gene expression, however, can be
challenging \cite{Elowitz:2002hb}. Naively, we could place Green
Fluorescent Protein (GFP) on a bacterial chromosome downstream of a
promoter that is activated by the system of interest. By measuring the
variation in fluorescence across a population of cells, we could
quantify stochasticity. Every biochemical reaction, however, is
potentially stochastic. Fluorescence variation could be because of
stochasticity in the process under study or could result from the
general background `hum' of stochasticity: stochastic effects in
ribosome synthesis could lead to different numbers of ribosomes and so
to differences in gene expression in each cell; stochastic effects in
the cell cycle machinery may desynchronize the population; stochastic
effects in signaling networks could cause each cell to respond
uniquely, and so on.

Variation has then two classes: {\bf intrinsic stochasticity}, the
stochasticity inherent in the dynamics of the system and that arises
from fluctuations in the timing of individual reactions, and {\bf
  extrinsic stochasticity}, the stochasticity originating from
reactions of the system of interest with other stochastic systems in
the cell or its environment \cite{Swain:2002kn,Elowitz:2002hb}. In
principle, intrinsic and extrinsic stochasticity can be measured by
creating a copy of the network of interest in the same cellular
environment as the original network \cite{Elowitz:2002hb}. We can
define intrinsic and extrinsic variables for the system of interest,
with fluctuations in these variables together generating intrinsic and
extrinsic stochasticity \cite{Swain:2002kn}. The intrinsic variables
of a system will typically specify the copy numbers of the molecular
components of the system. For gene expression, the level of occupancy
of the promoter by transcription factors, the numbers of mRNA
molecules, and the number of proteins are all intrinsic variables.
Imagining a second copy of the system -- an identical gene and
promoter elsewhere in the genome -- then the instantaneous values of
the intrinsic variables of this copy of the system will usually differ
from those of the original system. At any point in time, for example,
the number of mRNAs transcribed from the first copy of the gene will
usually be different from the number of mRNAs transcribed from the
second copy. Extrinsic variables, however, describe processes that
equally affect each copy of the system. Their values are therefore the
same for each copy. For example, the number of cytosolic RNA
polymerases is an extrinsic variable because the rate of gene
expression from both copies of the gene will increase if the number of
cytosolic RNA polymerases increases and decrease if the number of
cytosolic RNA polymerases decreases. In contrast, the number of
transcribing RNA polymerases is an intrinsic variable because we
expect the number of transcribing RNA polymerases to be different for
each copy of the gene at any point in time.

Stochasticity is quantified by measuring an intrinsic variable for
both copies of the system. For gene expression, the number of proteins
is typically measured by using fluorescent proteins as markers
\cite{Ozbudak:2002iq,Elowitz:2002hb,Blake:2003cn,Raser:2004gh}. Imaging
a population of cells then allows estimation of the distribution of
protein levels at steady-state. Fluctuations of the intrinsic variable
will {\it in vivo} have both intrinsic and extrinsic sources. The
number of proteins will fluctuate because of intrinsic stochasticity
generated during gene expression, but also because of stochasticity
in, for example, the number of cytosolic RNA polymerases or ribosomes
or proteosomes. We will use the term `noise' to mean an empirical
measure of stochasticity defined by the coefficient of variation (the
standard deviation divided by the mean) of a stochastic process. An
estimate of intrinsic stochasticity is the intrinsic noise which is
defined as a measure of the difference between the value of an
intrinsic variable for one copy of the system and its counterpart in
the second copy. For gene expression, typically the intrinsic noise is
the mean absolute difference (suitably normalized) at steady-state
between the number of proteins expressed from one copy of the gene and
the number of proteins expressed from the other copy
\cite{Elowitz:2002hb}. Such a definition supports the intuition that
intrinsic fluctuations cause variation in one copy of the system to be
uncorrelated with variation in the other copy. Extrinsic noise is
defined as the correlation coefficient between the intrinsic variable
of one copy of the system and its counterpart for the other copy
because extrinsic fluctuations equally affect both copies of the
system and consequently cause correlations between variation in one
copy and variation in the other. The intrinsic and extrinsic noise
should be related to the coefficient of variation of the intrinsic
variable of the original system of interest. This so-called total
noise is given by the square root of the sum of the squares of the
intrinsic and the extrinsic noise \cite{Swain:2002kn}.

Such two-colour measurements of stochasticity have been applied to
bacteria and yeast where gene expression has been characterized by
using two copies of a promoter placed in the genome with each copy
driving a distinguishable allele of Green Fluorescent Protein
\cite{Elowitz:2002hb,Raser:2004gh}. Both intrinsic and extrinsic noise
can be substantial giving, for example, a total noise of around 0.4,
and so the standard deviation of protein numbers is 40\% of the
mean. Extrinsic noise is usually higher than intrinsic noise. There
are some experimental caveats: both copies of the system should be
placed `equally' in the genome so that the probabilities of
transcription and replication are equal. This `equality' is perhaps
best met by placing the two genes adjacent to each other
\cite{Elowitz:2002hb}. Although conceptually there are no
difficulties, practically problems arise with feedback. If the protein
synthesized in one system can influence its own expression, the same
protein will also influence expression in a second copy of the
system. The two copies of the system have lost the (conditional)
independence they require to be two simultaneous measurements of the
same stochastic process.

\section*{A stochastic description of chemical reactions}

For any network of chemical reactions, the lowest level of description
commonly used in systems biology is the chemical master equation. This
equation assumes that the system is well-stirred and so ignores
spatial effects. It governs how the probability of the system being in
any particular state changes with time. A system state is defined by
the number of molecules present for each chemical species, and it will
change every time a reaction occurs. From the master equation we can
derive the deterministic approximation (a set of coupled differential
equations) which is often used to describe system dynamics. The
dynamics of the mean of each chemical species approximately obeys
these deterministic equations as the numbers of molecules of all
species increase \cite{Samoilov:2006ft,Grima:2010jl}. The master equation
itself is usually only solvable analytically for linear systems:
systems having only first-order chemical reactions.

Nevertheless, several approximations exist, all of which exploit the
tendency of fluctuations to decrease as the numbers of molecules
increase. The most systematic is the {\bf linear noise} approach of
van Kampen \cite{VanKampen:81:book}. If the concentration of each
chemical species is fixed, then changing the system volume, $\Omega$,
alters the number of molecules of every chemical species. The linear
noise approximation is based on a systematic expansion of the master
equation in the inverse of the system volume, $\Omega^{-1}$. It leads
to diffusion-like equations that accurately describe small
fluctuations around any stable attractor of the system. For systems
that tend to steady-state, a {\bf Langevin} approach is also often
used \cite{Gillespie:2000vv,Hasty:2000ku,Swain:2004ka}. Here
additive, white stochastic terms are included in the deterministic
equations, with the magnitude of these terms being determined by the
chemical reactions. At steady-state and for sufficiently high numbers
of molecules, the Langevin and linear noise approaches are equivalent.

Unfortunately, all these methods become intractable, in general, once
the number of chemical species in the system reaches more than three
(we then need to analytically calculate the inverse of at least a
$4\times4$ matrix or its eigenvalues). Rather than numerically solve
the master equation, the {\bf Gillespie algorithm}
\cite{Gillespie:1977ww}, a Monte Carlo method, is often used to
simulate intrinsic fluctuations by generating one sample time course
from the master equation. By doing many simulations and averaging, the
mean and variance for each chemical species can be calculated as a
function of time. Extrinsic fluctuations can be modelled as
fluctuations in the parameters of the system, such as the kinetic
rates \cite{Paulsson:2004dh,Shahrezaei:2008iq}. They can be included
by a minor modification of the Gillespie algorithm that feeds in a
pre-simulated time series of extrinsic fluctations and so generates
both intrinsic and extrinsic fluctuations \cite{Shahrezaei:2008iq}.

Here we will introduce the master equation and
briefly discuss the Gillespie algorithm.

\subsection*{The master equation}

Once molecules can react, the intrinsic stochsasticity destroys any
certainty of the numbers and types of molecules present, and we must
adopt a probabilistic description. For example, a model of gene
expression is given by
\reaction*{->[k] C ->[d] 0 }
where protein $C$ is synthesized on average every $1/k$ seconds and
degrades on average every $1/d$ seconds.  The reactions can be
described by the probability
$$
{\cal P}(\mbox{$n$ molecules of $C$ at time $t$})$$
and how this probability evolves with time. Each reaction rate is
interpreted as the probability per unit time of the appropriate
reaction.

We will write $P_n(t)$ for the probability that $n$ proteins exist at
time $t$ and consider the reactions that might have occurred just
prior to having $n$ molecules of protein. Let $\delta t$ be a time
interval small enough so that at most only one reaction can occur. If
there are $n$ proteins at time $t + \delta t$, then if a protein was
synthesized during the interval $\delta t$, there must have been $n-1$
proteins at time $t$. The probability of synthesis is
\equ
{\cal P}({\rm synthesis}) = k \delta t
\uqe
which is independent of the number of proteins present. If we have $n$
proteins at time $t+\delta t$ and a protein was degraded during the
interval $\delta t$, however, there must have been $n+1$ proteins at
time $t$. The probability of degradation is
\equ
{\cal P}({\rm degradation}) = (n+1) d \delta t .
\uqe
Neither synthesis nor degradation may have occurred during $\delta
t$. The number of proteins will be unchanged, which occurs with
probability
\equ
{\cal P}(\mbox{no reaction}) = 1 - k \delta t - n d \delta t .
\uqe
Notice that the probability of a protein degrading is $n d \delta t$
because $n$ proteins must have existed at time $t$.

Putting these probabilities together, we can the master equation
describing the time evolution of $P_n(t)$. Writing
\equn{poissmast} P_n(t + \delta t) = P_{n-1}(t) k \delta t +
P_{n+1}(t) d (n+1) \delta t + P_n(t) (1 - k \delta t - n d \delta t) .
\nuqe
dividing through by $\delta t$ and taking the limit $\delta t
\rightarrow 0$ gives
\equn{ms} 
\frac{\partial}{\partial t} P_n = k \Bigl[ P_{n-1} - P_n
\Bigr] - d \Bigl[ n P_n - (n+1) P_{n+1} \Bigr] 
\nuqe
Eq.\ \ref{ms} is an example of a master equation: all the
moments of the probability distribution $P_n(t)$ can be derived from
it.

Consider now a binary reaction:
\begin{equation}
\ce{A + B ->[f] C}
\label{nlreacs}
\end{equation}
where $A$ and $B$ bind irreversibly to form complex $C$ with
probability $f$ per unit time. Suppose further that individual $C$
molecules degrade with probability $d$ per unit time
\reaction*{C ->[d] 0}
The state of the system is then described by
$$
{\cal P}(\mbox{$n_A$ molecules of $A$, $n_B$ molecules of $B$, and
  $n_C$ molecules of $C$ at time $t$})
$$
which we will write as $P_{n_A,n_B,n_C}(t)$. We again consider a time
interval $\delta t$ small enough so that at most only one reaction can
occur. If the system at time $t+\delta t$ has $n_A$, $n_B$, and $n_C$
molecules of $A$, $B$, and $C$, then if reaction $f$ occurred during
the interval $\delta t$, the system must have been in the state
$n_A+1$, $n_B+1,$ and $n_C-1$ at time $t$. The probability of this
reaction is
\equ
{\cal P}(\mbox{$f$ reaction}) = f (n_A+1)(n_B+1) \delta t .
\uqe
Alternatively, reaction $d$ could have occurred during $\delta t$ and
so the system then must have been in the state $n_A$, $n_B$, and
$n_C+1$ at time $t$. Its probability is
\equ
{\cal P}(\mbox{$d$ reaction}) = d(n_C+1) \delta t .
\uqe
Finally, no reaction may have occurred at all, and so the system would
be unchanged at $t$ (in the state $n_A$, $n_B$, and $n_C$):
\equ
{\cal P}(\mbox{no reaction}) = 1-fn_An_B \delta t - d n_C \delta t .
\uqe

Thus we can find the master equation by writing
\eqan{m1}
\lefteqn{P_{n_A,n_B,n_C}(t+\delta t) =} &\nonu \\
& P_{n_A+1,n_B+1,n_C-1}(t) (n_A+1)(n_B+1)f \delta t 
 + P_{n_A,n_B,n_C+1}(t)(n_C+1)d \delta t \nonu \\
& + P_{n_A,n_B,n_C}(t)\Bigl[1-n_A n_B f \delta t - n_C d \delta t
\Bigr] \naqe
or
\eqan{m2}
\frac{\partial}{\partial t} P_{n_A, n_B, n_C} &=& f \Bigl[
(n_A+1)(n_B+1) P_{n_A+1,n_B+1,n_C-1} - n_A n_B P_{n_A,n_B,n_C} \Bigr] \nonu \\
& & - d \Bigl[ n_C P_{n_A,n_B,n_C} - (n_C+1) P_{n_A,n_B,n_C+1} \Bigr]
\naqe
in the limit of $\delta t \rightarrow 0$.

\subsection*{The definition of noise}
Noise is typically defined as the coefficient of variation: the ratio
of the standard deviation of a distribution to its mean. We will
denote noise by $\eta$:
\equn{noise}
\eta = \frac{\sqrt{\la N^2 \ra - \la N \ra^2}}{\la N \ra}
\nuqe
for a random variable $N$. The noise is dimensionless and measures the
magnitude of a typical fluctuation as a fraction of the mean.

\subsection*{Example: A birth-and-death processes}
The model of gene expression 
\begin{equation}
\ce{->[k] C ->[d] 0 } 
\label{bdreacs}
\end{equation}
is a birth-and-death process. Proteins can only be synthesized (born)
or degrade (die). We will solve the master equation for this system,
Eq.\ \ref{ms}, using a moment generating function.

The moment generating function for a probability distribution $P_n(t)$
is defined as
\equn{Fdef}
F(z,t)= \sum_{n=0}^\infty z^n P_n(t)
\nuqe
and can be thought of as a discrete transform. Differentiating the
moment generating function with respect to $z$ gives
\eqan{diffone}
\frac{\partial F}{\partial z} &=& \sum_{n=0}^\infty n z^{n-1} P_n  \label{difftwo} \\
\frac{\partial^2 F}{\partial z^2} &=& \sum_{n=0}^\infty n(n-1) z^{n-2} P_n . 
\naqe
The generating function and its derivatives have useful properties
because of their dependence on the probability distribution $P_n(t)$:
\eqan{prop1}
F(z=1, t) &=& \sum_{n=0}^\infty P_n(t) = 1 \\
\frac{\partial F}{\partial z}(z=1, t) &=& 
\sum_{n=0}^\infty n P_n(t) = \la n(t) \ra \label{prop2} \\
\frac{\partial^2 F}{\partial z^2}(z=1, t) &=& 
\sum_{n=0}^\infty n(n-1)P_n(t) = \la n^2(t) \ra - \la n(t) \ra . \label{prop3}
\naqe
Finding $F(z,t)$ therefore allows us to calculate all the moments of
$P_n(t)$: $F(z,t)$ is called the moment generating function.

The master equation can be converted into a partial differential
equation for the moment generating function. Multiplying (\ref{ms}) by
$z^n$ and summing over all $n$ gives
\eqa
\frac{\partial F}{\partial t} &=& k \sum_n z^n P_{n-1} - k F - d
\sum_n n z^n P_{n-1}
 + d \sum_n (n+1) z^n P_{n+1} \nonu \\
&=& k z \sum_n z^{n-1} P_{n-1} - k F - d z \sum_n n z^{n-1} P_n + d
\sum_n (n+1) z^n P_{n+1} \aqe
where we have factored $z$ out of some of the sums so that we can use
(\ref{Fdef}) and (\ref{diffone}). With these results and setting
$P_n=0$ if $n<0$, we can write
\equ \frac{\partial F}{\partial t} = k z F - F - d z \frac{\partial
  F}{\partial z} + d \frac{\partial F}{\partial z} \uqe
or
\equn{ms2}
\frac{\partial F}{\partial t} = (z-1) \left( k F - 
d \frac{\partial F}{\partial z} \right) .
\nuqe
This first order partial differential equation can be solved in
general using the method of characteristics \cite{VanKampen:81:book}.

We will solve (\ref{ms2}) to find the steady-state probability
distribution of protein numbers. At steady-state, $P_n(t)$ is
independent of time and so $\frac{\partial F}{\partial t} = 0$ from
(\ref{Fdef}). Consequently, (\ref{ms2}) becomes
\equ
\frac{\partial F}{\partial z} = \frac{k}{d} F 
\uqe
which is an ordinary differential equation. This equation has a
solution
\equ
F(z)= C \e^{\frac{k}{d} z}
\uqe
for some constant $C$. This constant can be determined from
(\ref{prop1}), implying
\equn{Ffinale}
F(z)= \e^{\frac{k}{d}(z-1)} .
\nuqe
By differentiation (\ref{Ffinale}) with respect to $z$ and using
(\ref{prop2}) and (\ref{prop3}), the moments of $n$ can be
calculated. For this case, we can Taylor expand (\ref{Ffinale}) and
find the probability distribution $P_n$ by comparing the expansion
with (\ref{Fdef}). Expanding gives
\equ
F(z)= \e^{-\frac{k}{d}} \sum_{n=0}^\infty \frac{ \left( k/d \right)^n}{n!} z^n
\uqe
implying that the steady-state probability of having $n$ proteins is
\equ
P_n= \e^{-k/d} \frac{\left(k/d \right)^n}{n!}
\uqe
which is a Poisson distribution. The first two moments are
\eqan{fish0}
\la n \ra &=& k/d \nonu \\
\la n^2 \ra - \la n \ra^2 &=& k/d \; = \; \la n \ra
\naqe
and consequently the noise is
\equn{fish}
\eta = 1/\sqrt{\la n \ra} 
\nuqe
from (\ref{noise}).

\begin{figure}[ht]
\begin{center}
\scalebox{0.4}{\includegraphics{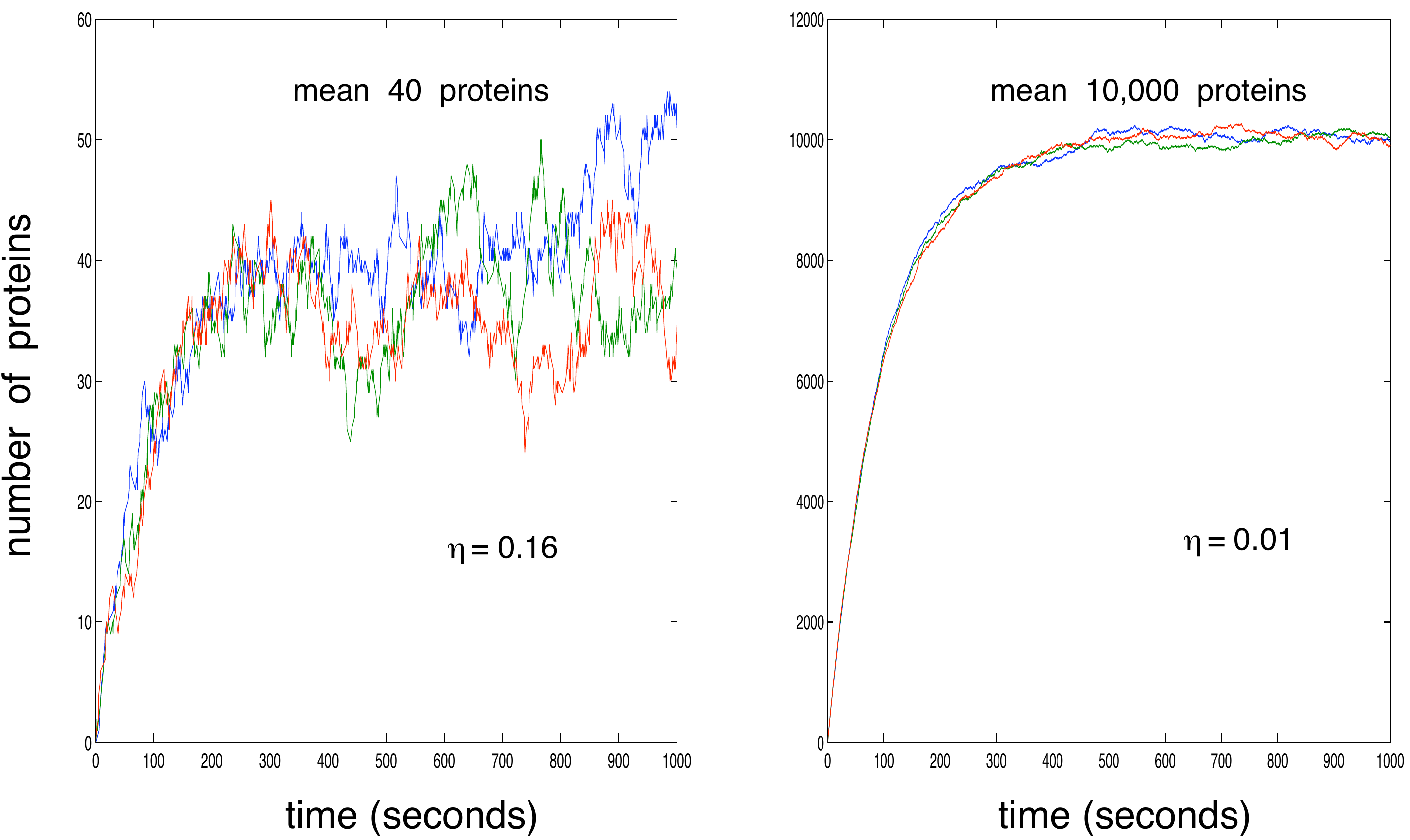}}
\caption{\small Three simulation runs of two birth-and-death models of
  gene expression (Eq.\ \ref{bdreacs}). Each model has different rate constants leading to different mean protein levels.} 
\label{fano}
\end{center}
\end{figure}

Eq.\ (\ref{fish}) demonstrates a `rule-of-thumb': stochasticity
generally become more significant as the number of molecules in the
system decrease (Fig.\ \ref{fano}). Approximate expression for the
distribution of proteins now exist for more realistic models of gene
expression \cite{Friedman:2006wa,Shahrezaei:2008bp}.

\subsubsection*{Recovering the deterministic equations}
Solving the master equation is possible for linear systems, i.e.\
those with only first-order chemical reactions, but often only at
steady-state \cite{VanKampen:81:book,Gardiner:90:book}. Solving for
the moments of a master equation is often easier.

For the non-linear system of Eq.\ \ref{nlreacs}, we will use the
master equation, (\ref{m2}), to derive the equation of motion for the
mean of $C$. The mean of $C$ is defined as
\equ
\la C(t) \ra = \sum_{n_A,n_B,n_C} n_C P_{n_A,n_B,n_C}(t)
\uqe
and is a function of time.

Multiplying (\ref{m2}) by $n_C$ and summing over $n_A$, $n_B$, and $n_C$ gives
\eqa
\frac{\partial}{\partial t} \la C \ra &=& f \sum  
(n_C - 1 + 1)(n_A+1)(n_B+1) P_{n_A+1,n_B+1,n_C-1} \nonu \\
& & - f \sum n_A n_B n_C P_{n_A,n_B,n_C}  - d \sum n_C^2 P_{n_A,n_B,n_C} \nonu \\
& & + d \sum (n_C+1-1)(n_C+1) P_{n_A,n_B,n_C+1} \aqe
where the terms in round brackets have been factored to follow the
subscripts of $P$. Therefore, by using results such as
\eqa
\la A B C \ra &=& \sum_{n_A,n_B,n_C=0}^\infty n_A n_B n_C P_{n_A,n_B,n_C} \nonu \\
&=& \sum_{n_A,n_B,n_C=0}^\infty (n_A+1)(n_B+1)(n_C-1) P_{n_A+1,n_B+1,n_C-1}
\aqe
as $P_{n_A,n_B,n_C}(t)$ is zero if any of $n_A$, $n_B$, or $n_C$ are
negative, we have
\eqan{mC}
\frac{\partial}{\partial t} \la C \ra &=& f \Bigl[ \la ABC \ra + \la AB \ra \Bigr] - f \la ABC \ra  -d \la C^2 \ra + d \Bigl[ \la C^2 \ra - \la C \ra \Bigr] \nonu \\
&=& f \la A B\ra - d\la C \ra
\naqe
which is the microscope equation for the dynamics of the mean of $C$.

We can also consider the deterministic equation for the
dynamics. Applying the law of mass action to this system, the
concentration of $C$, $[C]$, obeys
\equn{deterministic}
\frac{d}{dt} [C] = \tilde{f} [A][B] - \tilde{d} [C]
\nuqe
where $\tilde{f}$ and $\tilde{d}$ are the macroscopic (deterministic)
rate constants. The macroscopic concentration is related to the mean
number of molecules by
\equn{conc}
[C]= \frac{\la C \ra}{V}
\nuqe
and so the deterministic equations are equations for the rate of
change of the means of the different chemical species: using
(\ref{conc}), (\ref{deterministic}) becomes
\equn{det}
\frac{d}{dt} \la C \ra = \frac{\tilde{f}}{V} \la A \ra \la B\ra - \tilde{d} \la C \ra .
\nuqe

By comparing the deterministic equation, (\ref{det}), with the
microscopic equation, (\ref{mC}), we can relate the stochastic
probabilities of reaction per unit time and the deterministic kinetic
rates:
\eqan{conv1}
\tilde{f} &=& \frac{V \la A B \ra}{\la A \ra \la B \ra} \cdot f  \nonu \\
\tilde{d} &=& d 
\naqe
For first-order reactions both the kinetic rate and the probability
are the same. The macroscopic rate $\tilde{f}$ is usually measured
under conditions where the deterministic approximation holds and
numbers of molecules are large. We can write
\eqan{conv2}
\tilde{f} &=& \frac{V \Bigl(\la A \ra \la B \ra + \la A B \ra - \la A \ra \la B \ra \Bigr)}{\la A \ra \la B \ra} \cdot f \nonu \\
&=& V f \cdot \left( 1 + \frac{\la A B \ra - \la A \ra \la B \ra}{\la A \ra \la B \ra} \right) \nonu \\
& \simeq & V f
\naqe
where the fluctuation term becomes negligible as the numbers of
molecules increase because its numerator, the co-variance $\la A B \ra
- \la A \ra \la B \ra$, is expected to be proportional to the mean
number of molecules, while its denominator is proportional to the
square of the mean number of molecules. Eq.\ (\ref{fish0}) is an
explicit example of this statement. Eq.\ (\ref{conv2}) is almost
always used to relate the macroscopic rate and the probability of
reaction for second-order reactions.

\subsubsection*{An exception: homo-dimerization reactions}
A homo-dimerization reaction 
\reaction*{A + A ->[f] A_2}
occurs when two identical monomers combine to form a dimer. This
reaction is common among transcription factors. The master
equation is now
\equ
\frac{\partial P_{n_A}}{\partial t} = f \left[
  \left( \begin{array}{c} n_A+2 \\ 2 \end{array} \right) P_{n_A+2} -
  \left( \begin{array}{c} n_A \\ 2 \end{array} \right) P_{n_A} \right]
\uqe
where each coefficient is the number of ways of forming a dimer. Eq.\
(\ref{conv1}) becomes
\equ
2 \frac{\tilde{f}}{V} \la A \ra^2 = f \la A (A-1) \ra .
\uqe
Assuming that $\tilde{f}$ is measured for large numbers of molecules,
we can write
\equ
\la A(A-1) \ra \simeq \la A \ra^2
\uqe
and so to
\equ
\tilde{f} \simeq \frac{f V}{2}
\uqe
which is the inter-conversion formula for dimerization reactions.

\section*{Simulating stochastic biochemical reactions}
The Gillespie algorithm \cite{Gillespie:1977ww} is most commonly used
to simulate intrinsic fluctuations in biochemical systems. The
equivalent of two dice are rolled on the computer: one to choose which
reaction will occur next and the other to choose when that reaction
will occur. Assume that we have a system in which $n$ different
reactions are possible, then the probability that starting from time
$t$ a reaction only occurs between $t+\tau$ and $t+\tau+\delta \tau$
must be calculated for each reaction. Let this probability be
$P_i(\tau)\delta \tau$ for reaction $i$, say.

For example, if reaction $i$ corresponds to the second-order reaction
of Eq.\ \ref{nlreacs}, then
\eqan{prop}
{\cal P}(\mbox{reaction $i$ in time $\delta \tau$}) &=& n_A n_B f \delta \tau \nonu \\
& = & a_i \delta \tau
\naqe
where $a_i$ is referred to as the propensity of reaction
$i$. Therefore,
\eqan{prob}
P_i(\tau)\delta \tau &=& {\cal P}(\mbox{no reaction for time $\tau$}) \nonu \\
& & \times {\cal P}(\mbox{reaction $i$ happens in time $\delta \tau$}) \nonu \\
& \equiv & P_0(\tau) a_i \delta \tau
\naqe
with $P_0(\tau)$ the probability that no reaction occurs during the
interval $\tau$. This probability is the product of the probability of
having no reactions at time $\tau$ and the probability of no reactions
occurring in time $\delta \tau$:
\equ
P_0(\tau+\delta \tau) = P_0(\tau) \Bigl[ 1 - \sum_{j=1}^n a_j \delta \tau \Bigr]
\uqe
which implies
\equ
\frac{dP_0}{d\tau} = - P_0 \sum^n_{j=1} a_j 
\uqe
and so
\equn{p0}
P_0(\tau) = \exp \left( - \tau \sum a_j \right) .
\nuqe
Thus we have
\equ
P_i(\tau) = a_i \e^{-\tau \sum a_j}
\uqe
from (\ref{p0}). 

To choose which reaction to simulate, an $n$-sided die is rolled with
each side corresponding to a reaction and weighted by the reaction's
propensity. A second die is then used to determine the time when the
reaction occurs by sampling from (\ref{p0}). All the chemical species
and the time variable are updated to reflect the occurrence of the
reaction, and the process is then repeated. See Gillespie (1977)
\cite{Gillespie:1977ww} for more details.

\begin{figure}[ht]
\begin{center}
\scalebox{0.5}{\includegraphics{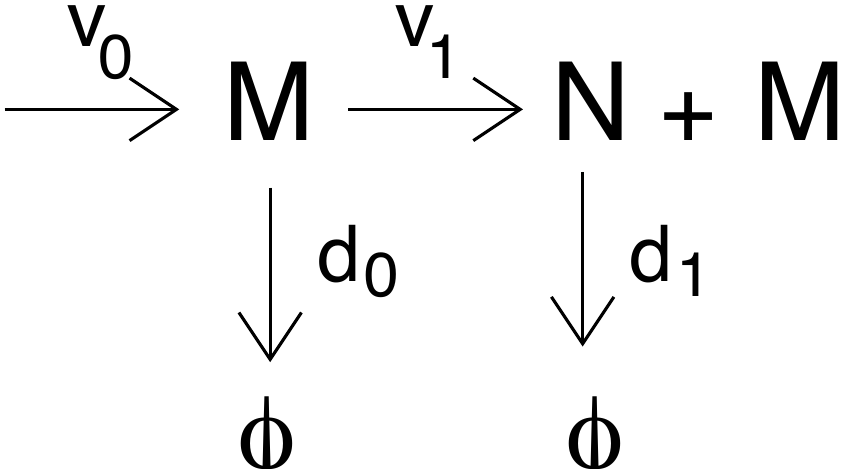}}
\caption{\small A model of gene expression that explicitly includes
  transcription (rate $v_0$) and translation (rate $v_1$) as
  first-order processes. mRNA is denoted by $M$ and protein by $N$.}
\label{ge}
\end{center}
\end{figure}

Extrinsic fluctuations can be included by considering reaction rates
that change with time \cite{Shahrezaei:2008iq}. A reaction rate is often a
function of the concentration of another protein and so fluctuates
because this protein concentration fluctuates. For example, $v_0$ in
Fig.\ \ref{ge} is a function of the concentration of free RNA
polymerases and $v_1$ is a function of the concentration of free
ribosomes. By simulating extrinsic fluctuations with the desired
properties before running the Gillepsie algorithm and then
approximating this extrinsic time series by a sequence of linear
changes over small time intervals, we can `feed' the extrinsic
fluctuations into the Gillepsie algorithm and so let a parameter, or
many parameters, fluctuate extrinsically.

\section*{Langevin theory: an improved model of gene expression}

We can model transcription and translation as first-order reactions \cite{Thattai:2001hj}. Both mRNA, $M$, and protein, $N$, are present,
and each has their own half-life (determined by the inverse of their
degradation rates).

\subsection*{The Langevin solution}
Langevin theory gives an approximation to the solution of the master
equation. It is strictly only valid when numbers of molecules are
large. Stochastic terms are explicitly added to the deterministic
equations of the system. For the model of Fig.\ \ref{ge}, the
deterministic equations are
\eqa
\frac{dM}{dt} &=& v_0- d_0 M \nonu \\
\frac{dN}{dt} &=& v_1 M - d_1 N .
\aqe
A Langevin model adds a stochastic variable, $\xi(t)$, to each
\eqan{addnoise}
\frac{dM}{dt} &=& v_0- d_0 M + \xi_1(t) \nonu \\
\frac{dN}{dt} &=& v_1 M - d_1 N + \xi_2(t)
\naqe
and is only fully specified when the probability distributions for the
$\xi_i$ are given. The $\xi_i$ must be specified so that they mimic
thermal fluctuations and model intrinsic fluctuations. The solution of
the Langevin equation should then be a good approximation to that of
the Master equation (and an exact solution in some limit).

To define $\xi$, we must give its mean and variance as functions of
time and its autocorrelation.

\subsubsection*{Understanding stochasticity: autocorrelations}
\begin{figure}[ht]
\begin{center}
\scalebox{0.5}{\includegraphics{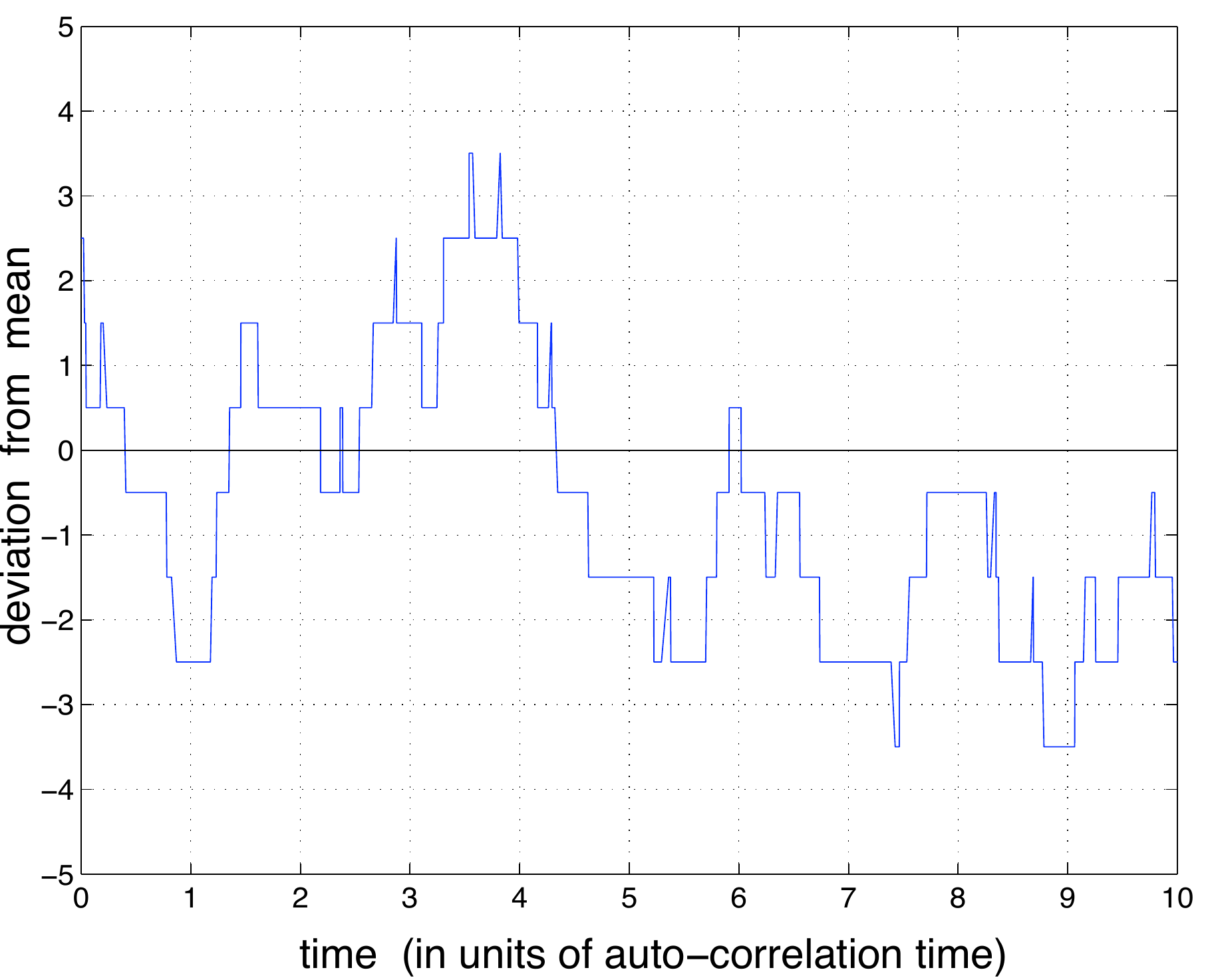}}
\caption{\small A time-series of a birth-death process. Time has been rescaled by the autocorrelation
  time. The deviation from the mean, $n - \la n \ra$, in numbers of
  molecules is plotted on the $y$-axis.}
\label{auto0}
\end{center}
\end{figure}

The autocorrelation time of a stochastic variable describes the
average life-time of a typical fluctuation. We will denote it by
$\tau$. Fig.\ \ref{auto0} shows typical behaviour of a stochastic
variable obeying a Poisson distribution. Time has been rescaled by the
autocorrelation time. On average, the number of molecules changes
significantly only over a time $\tau$ (1 in these units).

The autocorrelation time is found from the autocorrelation
function. For a stochastic variable $N$, the autocorrelation function
is
\eqan{corr}
C_N(t_1,t_2) &=& \left \la \Bigl[ N(t_1) - \la N(t_1) \ra \Bigr] \Bigl[ N(t_2) - \la N(t_2) \ra \Bigr] \right \ra  \nonu \\
&=& \left\la \Bigl \{ N(t_1) N(t_2) - \la N(t_1) \ra N(t_2) - N(t_1) \la N(t_2) \ra + \la N(t_1) \ra \la N(t_2) \ra \Bigr \} \right\ra \nonu \\
&=& \la N(t_1) N(t_2) \ra - \la N(t_1) \ra \la N(t_2) \ra .
\naqe
It quantifies how a deviation of $N$ away from its mean at time $t_1$
is correlated with the deviation from the mean at a later time
$t_2$. It is determined by the typical life-time of a
fluctuation. When $t_1=t_2$, (\ref{corr}) is just the variance of
$N(t)$.

\begin{figure}[ht]
\begin{center}
\scalebox{0.5}{\includegraphics{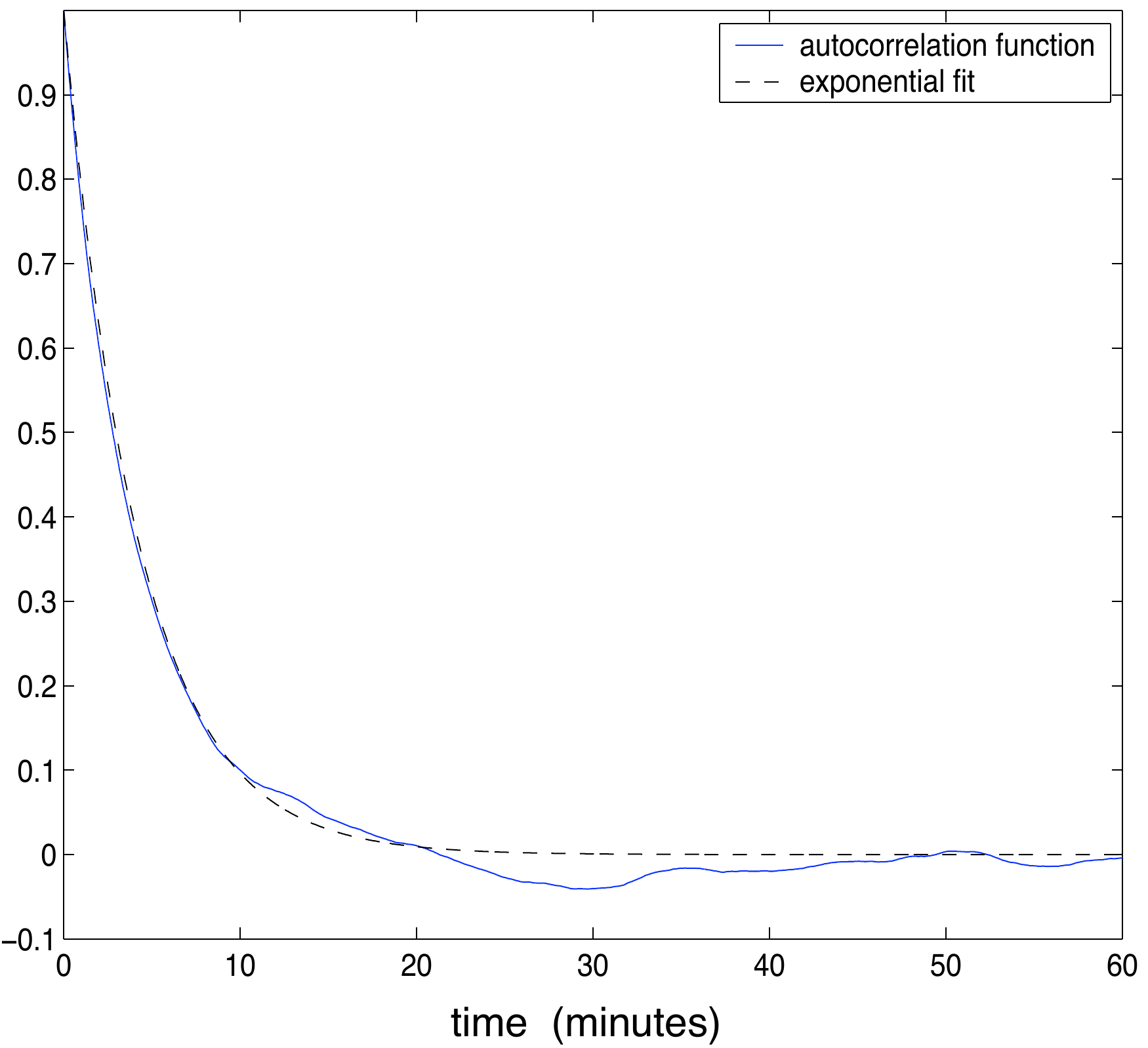}}
\caption{\small Auto-correlation function for a birth-death process. The dotted line is an exponential fit using an autocorrelation time of $1/d \simeq 4.2 \; {\rm minutes}$.}
\label{auto1}
\end{center}
\end{figure}

Stationary processes are processes that are invariant under time
translations and so are statistically identical at all time
points. For a stationary process, such as the steady-state behaviour
of a chemical system, the autocorrelation function obeys
\equ
C_N(t_1,t_2) = C_N(t_2-t_1) .
\uqe
It is a function of one variable: the time difference between the two
time points considered. Fig.\ \ref{auto1} shows the steady-state
autocorrelation function for the Poisson model of gene expression. It
is normalized by the variance and is fit well by an exponential decay:
$\e^{-t/\tau}$. A typical fluctuation only persists for the timescale
$\tau$ as enough new reaction events occur during $\tau$ to
significantly change the dynamics and remove any memory the system may
have had of earlier behaviour.

For linear systems, the time-scale associated with degradation
determines the steady-state autocorrelation time. Degradation provides
the restoring force that keeps the number of proteins fluctuating
around their mean steady-state value. The probability of degradation
in time $\delta t$, $d \times n \times \delta t$, changes as the
number of proteins $n$ changes. It increases as the number of proteins
rises above the mean value, increasing the probability of degradation
and of return to mean levels; it decreases as the number of proteins
falls below mean levels, decreasing the probability of degradation and
increasing again the probability of returning to mean values. For a
linear system with multiple time-scales, the autocorrelation function
is a sum of terms, each exponentially decreasing with $t_1-t_2$ at a
time-scale set by the inverse of a degradation-like rate.

\subsubsection*{White noise}
In Langevin theory, a stochastic variable, $\xi$, is added to each
deterministic equation. This variable describes thermal fluctuations:
those fluctuations that arise from collisions of the molecule of
interest with surrounding molecules. Such collisions act to either
increase or decrease the probability of reaction. {\it A priori},
there is no reason why thermal fluctuations would favour one effect
over the other and so $\xi(t)$ is defined to have a mean of zero:
\equn{wn1}
\la \xi(t) \ra = 0 .
\nuqe

The time-scale associated with collisions is assumed to be much
shorter than the time-scale of a typical reaction. The changes in
internal energy and position of the molecule of interest because of
collisions with solvent molecules are therefore uncorrelated at the
reaction time-scale. Mathematically, the autocorrelation time, $\tau$,
of the autocorrelation function
\equ
C_\xi(t_2-t_1) = \la \xi(t_1) \xi(t_2) \ra
\uqe
is taken to zero. If $\Gamma/\tau$ is the variance of $\xi$ at time
$t$, the auto-correlation function is
\equ
C_\xi(t_2-t_1) = \frac{\Gamma}{\tau} \e^{-(t_2-t_1)/\tau}
\uqe
which becomes
\equn{wn2}
\la \xi(t_1) \xi(t_2) \ra = \Gamma \delta(t_2-t_1)
\nuqe
in the limit of $\tau \rightarrow 0$ where $\delta(t)$ is the Dirac
delta function. A stochastic variable that obeys (\ref{wn1}) and
(\ref{wn2}) is referred to as `white'. It is completely uncorrelated
in time and has zero mean. Stochastic variables with zero mean and a
finite auto-correlation time are considered `coloured'. The parameter
$\Gamma$ determines the magnitude of fluctuations and needs to be
carefully specified (see \cite{VanKampen:81:book} for a discussion of
how Einstein famously chose $\Gamma$ to appropriately model Brownian
motion).

\subsubsection*{Langevin theory for stochastic gene expression}
We now return to modelling the gene expression of Fig.\ \ref{ge}. Eq.\
(\ref{addnoise}) is shown again below
\eqan{ge2}
\frac{dM}{dt} &=& v_0- d_0 M + \xi_1(t) \nonu \\
\frac{dN}{dt} &=& v_1 M - d_1 N + \xi_2(t)
\naqe
and is the deterministic equations of Fig.\ \ref{ge} with additive,
white stochastic variables.

Although we expect $\xi_1$ and $\xi_2$ to have zero mean and zero
autocorrelation times, we can show that this assumptions are true
explicitly by first considering the steady-state solution of
(\ref{ge2}) in the absence of the stochastic variables $\xi_i$:
\eqan{ss}
M_s = \frac{v_0}{d_0} &;& N_s = \frac{v_1}{d_1} M_s
\naqe
If we assume that the system is at or very close to steady-state, and
consider a time interval $\delta t$ small enough such that at most
only one reaction can occur, then $\xi_1$ and $\xi_2$ can only have
the values
\equ
\xi_i \delta t = \left\{ \begin{array}{l} +1 \\ 0 \\ -1 \end{array} \right.
\uqe
where $i=1$ or 2, as the number of $N$ or $M$ molecules can only
increase or decrease by one or remain unchanged in time $\delta t$.

Define
$$
P(i,j)= {\cal P}(\xi_1 \delta t=i, \xi_2 \delta t= j)
$$  
i.e.\ the probability that the number of mRNAs changes by an amount $i$
and that the number of proteins changes by an amount $j$. Then the
reaction scheme of Fig.\ \ref{ge} implies
\eqan{probs}
P(+1,0) &=& v_0 \delta t \nonu \\
P(+1,-1) &=& 0 \nonu \\
P(+1,+1) &=& 0 \nonu \\
& & \nonu \\
P(-1,0) &=& d_0 M_s \delta t \nonu \\
P(-1,+1) &=& 0 \nonu \\
P(-1,-1) &=& 0 \nonu \\
& & \nonu \\
P(0,+1) &=& v_1 M_s \delta t \nonu \\
P(0,0) &=& 1 - v_0 \delta t -v_1 M_s \delta t - d_0 M_s \delta t - d_1 N_s \delta t\nonu \\
P(0,-1) &=& d_1 N_s \delta t 
\naqe
at steady-state. 

We can use these probabilities to calculate the moments of the
$\xi_i$. First,
\eqan{me1}
\la \xi_1 \delta t \ra &=& (+1)\times v_0 \delta t + (-1)\times d_0 M_s \delta t + (0)\times(1- v_0 \delta t- d_0 M_s \delta t) \nonu \\
&=& (v_0 - d_0 M_s) \delta t \nonu \\
&=& 0
\naqe
and
\eqan{me2}
\la \xi_2 \delta t \ra &=& (+1)\times v_1 M_s \delta t + (-1)\times d_1 N_s \delta t \nonu \\
&=& (v_1 M_s - d_1 N_s) \delta t \nonu \\
&=& 0
\naqe
using (\ref{ss}). The means are both zero, as expected, and the
$\xi_i$ act to keep the system at steady-state (as they should).

For the mean square, we have
\eqan{x1}
\la \xi_1^2 \delta t^2 \ra &=& (+1)^2 \times v_0 \delta t + (-1)^2 \times d_0 M_s \delta t \nonu \\
&=& (v_0 + d_0 M_s) \delta t \nonu \\
&=& 2 d_0 M_s \delta t
\naqe
or
\equ
\la \xi_1^2 \ra = \frac{2 d_0 M_s}{\delta t}
\uqe
and, similarly,
\eqan{x2}
\la \xi_2^2 \ra &=& \frac{2 d_1 N_s}{\delta t} \nonu \\
\la \xi_1 \xi_2 \ra &=& 0
\naqe

If the system is close to steady-state and the steady-state values of
$M_s$ and $N_s$ are large enough such that
\eqan{ss2}
|M-M_s| \ll M_s &;& |N-N_s| \ll N_s
\naqe
then we can assume that (\ref{probs}) is valid for all
times. Consequently, $\xi_1$ at time $t_2$, say, is completely
uncorrelated with $\xi_1$ at time $t_1$, where $|t_2-t_1| > \delta t$
(just as the throws of a die whose outcomes are also given by fixed
probabilities and are uncorrelated). Thus, we define as white
stochastic terms
\eqan{noises}
\la \xi_1(t_1) \xi_1(t_2) \ra &=& 2 d_0 M_s \delta(t_2-t_1) \nonu \\
\la \xi_2(t_1) \xi_2(t_2) \ra &=& 2 d_1 N_s \delta(t_2-t_1) \nonu \\
\la \xi_1(t_1) \xi_2(t_2) \ra &=& 0
\naqe
with their magnitudes coming from (\ref{x1}) and (\ref{x2}).

This definition of $\xi_1$ and $\xi_2$ implies that the steady-state
solution of (\ref{ge2}) will have the true mean and variance of $N$
and $M$ obtained from the master equation, providing (\ref{ss2}) is
obeyed.

\subsubsection*{A further simplification}
\begin{figure}[ht]
\begin{center}
\scalebox{0.6}{\includegraphics{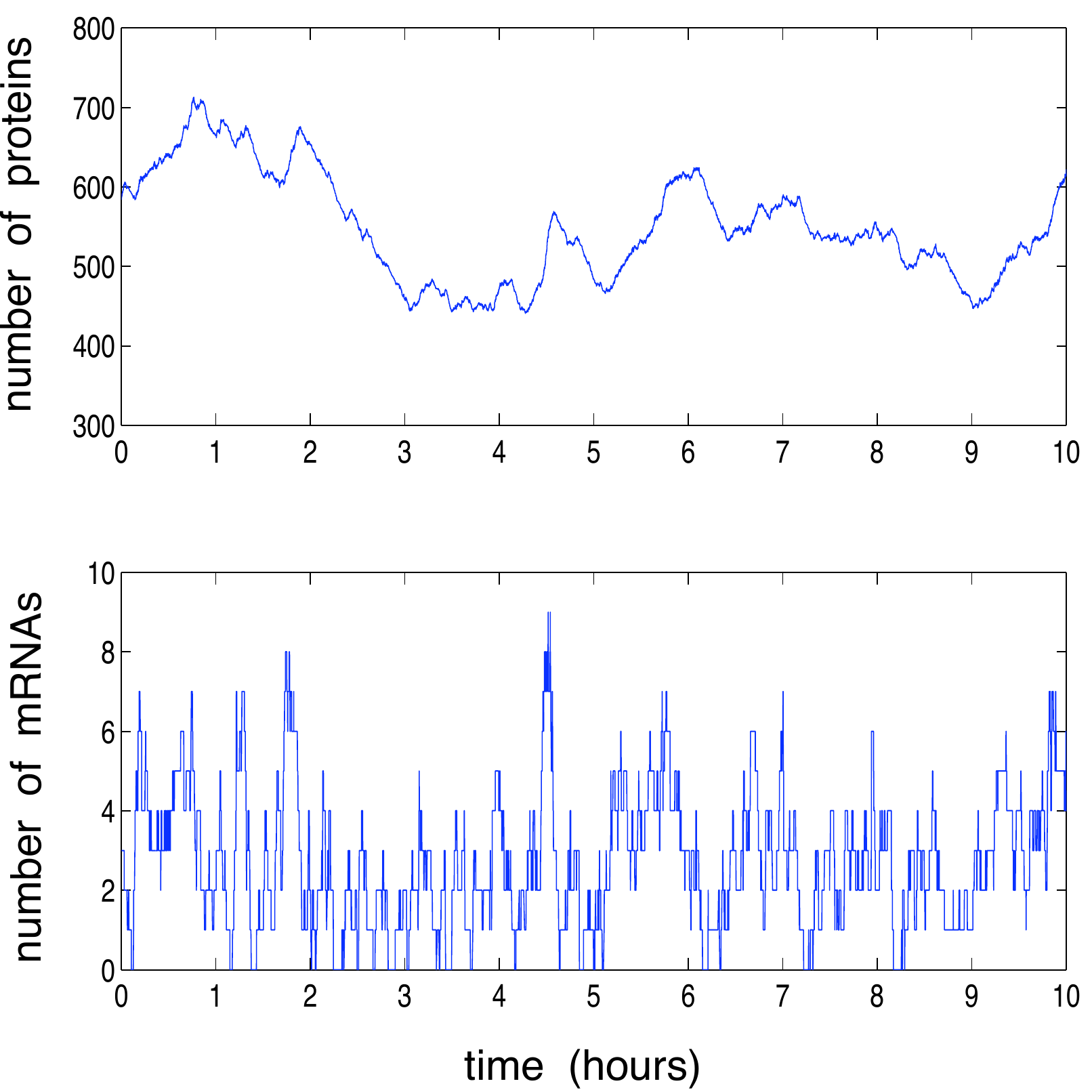}}
\caption{\small Protein and mRNA numbers from a simulation of the scheme of Fig.\ \ref{ge}. Protein half-life is approximately 1 hour while that of mRNA is only 3 minutes.} 
\label{MN}
\end{center}
\end{figure}

Although it is possible to directly solve the two coupled differential
equations of (\ref{ge2}), we can also take advantage of the different
time-scales associated with mRNA and protein. Typically, mRNA
life-time is of order minutes while protein life-time is of order
hours in bacteria. Fig.\ \ref{MN} shows a simulated time series of
protein and mRNA: protein has a longer autocorrelation time of $1/d_1$
compared to the mRNA autocorrelation time of $1/d_0$.

Many mRNA fluctuations occur during one protein fluctuation, and so
the mean level of mRNA reaches steady-state relatively
quickly. Therefore, we can set
\equ
\frac{dM}{dt} \simeq 0 
\uqe
which implies that
\eqa
M & = & \frac{v_0}{d_0} + \frac{\xi_1}{d_0} \nonu \\
&=& M_s + \frac{\xi_1}{d_0}
\aqe
Consequently, the equation for protein, (\ref{ge2}), becomes
\equn{N}
\frac{dN}{dt} = v_1 M_s - d_1 N + \frac{v_1}{d_0} \xi_1 + \xi_2
\nuqe
and so is a function of the two stochastic variables $\xi_1$ and
$\xi_2$. To simplify (\ref{N}), we define a new stochastic variable
\equ
\Psi =  \frac{v_1}{d_0} \xi_1 + \xi_2
\uqe
which has mean
\equn{Psi0}
\la \Psi \ra =  \frac{v_1}{d_0} \la \xi_1 \ra + \la \xi_2 \ra = 0
\nuqe
from (\ref{me1}) and (\ref{me2}), and mean square
\eqa
\la \Psi(t_1) \Psi(t_2) \ra &=& \left(\frac{v_1}{d_0}\right)^2 \la \xi_1(t_1) \xi_1(t_2) \ra +  2 \left(\frac{v_1}{d_0}\right) \la \xi_1(t_1) \xi_2(t_2) \ra \nonu \\
& & + \la \xi_2(t_1) \xi_2(t_2) \ra 
\aqe
From Eqs.\ (\ref{noises}), this result simplifies
\eqan{psi}
\la \Psi(t_1) \Psi(t_2) \ra &=&  \left(\frac{v_1}{d_0}\right)^2 2d_0 M_s \delta(t_2-t_1) + 2 d_1 N_s \delta(t_2-t_1) \nonu \\
&=& 2 \left[ \frac{v_1^2}{d_0} M_s + d_1 N_s \right] \delta(t_2-t_1) \nonu \\
&=& 2 d_1 \left[ \frac{v_1}{d_1} M_s \frac{v_1}{d_0} + N_s \right] \delta(t_2-t_1) \nonu \\
&=& 2 d_1 N_s \left[ 1 + \frac{v_1}{d_0} \right] \delta(t_2-t_1)
\naqe
and so we need only consider one equation: 
\equn{N2}
\frac{dN}{dt} = v_1 M_s - d_1 N + \Psi(t)
\nuqe
The effects of the mRNA fluctuations have been absorbed into the
protein fluctuations and their magnitude has increased: compare
(\ref{noises}) and (\ref{psi}).

\subsubsection*{Solving the model}
Eq.\ (\ref{N2}) can be written as
\equ
\frac{d}{dt} \left( N \e^{d_1 t} \right) = v_1 M_s \e^{d_1 t} + \Psi \e^{d_1 t} 
\uqe
and so integrated
\equ
N(t) \e^{d_1 t} - N_s = \frac{v_1 M_s}{d_1} \left( \e^{d_1 t} - 1 \right) + \int_0^t \Psi(t') \e^{d_1 t'} dt'
\uqe
where we have assumed that $N= N_s$ when $t=0$. Thus
\equn{Nsol}
N(t) = N_s + \e^{-d_1 t} \int_0^t \Psi(t') \e^{d_1 t'} dt'
\nuqe

Using the properties of $\Psi(t)$, (\ref{Psi0}) and (\ref{psi}), as
well as (\ref{Nsol}), the mean protein number satisfies
\eqa
\la N(t) \ra &=& N_s + \e^{-d_1 t} \int_0^t \la \Psi(t') \ra \e^{d_1 t'} dt' \nonu \\
&=& N_s
\aqe
and so the steady-state is stable to fluctuations (as expected). 

We can also use (\ref{Nsol}) to find the autocorrelation function of
the protein number:
\eqa
\lefteqn{\la N(t_1) N(t_2) \ra } \nonu \\
& &=  \biggl \la \left[ N_s + \e^{-d_1 t_1} \int_0^{t_1} \Psi(t') \e^{d_1 t'} dt' \right] \times \left[ N_s + \e^{-d_1 t_2} \int_0^{t_2} \Psi(t'') \e^{d_1 t''} dt'' \right] \biggr \ra \nonu \\
& & = N_s^2 + \e^{-d_1(t_1+t_2)} \int_0^{t_1} \e^{d_1 t'} dt' \int_0^{t_2} \e^{d_1 t''} dt'' \la \Psi(t') \Psi(t'') \ra
\aqe
as $\la \Psi \ra= 0$. From (\ref{psi}), we then have
\equn{n20}
\la N(t_1) N(t_2) \ra - N_s^2 = 2 d_1 N_s \left( 1 + \frac{v_1}{d_0} \right) \e^{-d_1(t_1+t_2)} \int_0^{t_1} dt' \int_0^{t_2} dt'' \e^{d_1(t'+t'')} \delta(t'-t'')
\nuqe
To evaluate the double integral, we need to determine when $t'$ is
equal to $t''$. If $t_2 \ge t_1$, then the integral can be decomposed
into
\eqa
\int_0^{t_2} dt' \int_0^{t_1} dt'' &=& \left( \int_{t_1}^{t_2} dt' + \int_0^{t_1} dt' \right) \int_0^{t_1} dt'' \nonu \\
&=& \int_{t_1}^{t_2} dt' \int_0^{t_1} dt'' + \int_0^{t_1} dt'  \int_0^{t_1} dt''
\aqe
where we now explicitly see that $t' > t''$ for the first term (and
there will be no contribution from the delta function) and $t'$ can
equal $t''$ for the second term (and there will be a contribution from
the delta function). Therefore,
\eqa
\lefteqn{\int_0^{t_2} dt' \int_0^{t_1} dt'' \e^{d_1(t'+t'')} \delta(t'-t'') } \nonu \\
& & = \int_{t_1}^{t_2} dt' \int_0^{t_1} dt'' \e^{d_1(t'+t'')} \delta(t'-t'') + \int_0^{t_1} dt'  \int_0^{t_1} dt''   \e^{d_1(t'+t'')} \delta(t'-t'') \nonu \\
& & = \int_0^{t_1} \e^{2 d_1 t'} dt' \nonu \\
& & = \frac{1}{2d_1} \left( \e^{2 d_1 t_1} -1 \right)
\aqe
because the first integral evaluates to zero.

Consequently, (\ref{n20}) becomes
\eqa
\la N(t_1) N(t_2) \ra - N_s^2 &=& 2 d_1 N_s \left( 1 + \frac{v_1}{d_0} \right) \e^{-d_1(t_1 + t_2)} \frac{1}{2d_1} \left( \e^{2 d_1 t_1} - 1 \right) \nonu \\
&=& N_s \left( 1 + \frac{v_1}{d_0} \right) \left( \e^{-d_1(t_2-t_1)} - \e^{-d_1(t_1+t_2)} \right)
\aqe
and we finally have
\equn{corrN}
\la N(t_1) N(t_2) \ra - \la N(t_1)\ra \la  N(t_2) \ra = N_s  \left( 1 + \frac{v_1}{d_0} \right) \left( \e^{-d_1(t_2-t_1)} - \e^{-d_1(t_1+t_2)} \right)
\nuqe
as $\la N(t) \ra = N_s$. Eq.\ (\ref{corrN}) is the autocorrelation
function for protein number and becomes
\equn{cN}
C_N = N_s \left(1 + \frac{v_1}{d_0} \right) \e^{-d_1(t_2-t_1)}
\nuqe
after long times $t_2 > t_1 \gg 1$. The protein autocorrelation time
is $1/d_1$.

We can also find similar expressions for mRNA. Eq.\ (\ref{N2}) has the
same structure as the equation for mRNA
\equn{M2}
\frac{dM}{dt} = v_0- d_0 M + \xi_1(t) 
\nuqe
with a constant rate of production and first-order degradation. The
solution of (\ref{M2}) will therefore be of the same form as
(\ref{cN}), but with $d_1$ replaced by $d_0$ and the magnitude of the
stochastic term coming from (\ref{noises}) rather than
(\ref{psi}). This substitution gives
\equn{cM}
C_M = M_s  \e^{-d_0(t_2-t_1)}
\nuqe
so that the autocorrelation time of the mRNA is $1/d_0$.

We can calculate the noise in mRNA when $t_1 = t_2$ because then the
autocorrelation becomes the variance:
\eqan{etaM}
\eta^2_M &=& \frac{\la M(t)^2 \ra - \la M(t) \ra^2}{\la M(t) \ra^2} \nonu \\
&=& \frac{M_s}{M_s^2} \nonu \\
&=& \frac{1}{\la M \ra}
\naqe
Eqs.\ (\ref{cM}) and (\ref{etaM}) are the solutions to any
birth-and-death model and correspond to the expressions given in
(\ref{fish0}) and (\ref{fish}).

The protein noise is a little more complicated. It satisfies
\eqan{etaN}
\eta^2_N &=& \frac{1}{N_s} + \frac{v_1}{d_0} \frac{1}{N_s} \nonu \\
&=& \frac{1}{N_s} + \frac{d_1}{d_0} \frac{1}{M_s} \nonu \\
&=& \frac{1}{\la N \ra} +  \frac{d_1}{d_0} \frac{1}{\la M \ra}
\naqe
which should be compared with (\ref{fish}) for a birth-death process. The mRNA acts as a fluctuating source of proteins and
increases the noise above the Poisson value. Eq.\ (\ref{etaN}) can be
described as
\equ
(\mbox{protein noise})^2 = (\mbox{Poisson noise})^2 + \frac{\mbox{mRNA lifetime}}{\mbox{protein lifetime}} \times (\mbox{mRNA noise})^2
\uqe
The Poisson noise is augmented by a time average of the mRNA noise. As
the protein life-time increases compared to the mRNA life-time, each
protein averages over more mRNA fluctuations and the overall protein
noise decreases. Ultimately, $\eta_N$ approaches the Poisson result as
$d_1/d_0 \rightarrow 0$.

More generally, we should include active and inactive states of the
promoter. With this extension, the model of gene expression appears
valid for bacteria \cite{Golding:2005dm}, yeast \cite{Raser:2004gh},
slime moulds \cite{Chubb:2006kz}, and mammalian cells \cite{Raj:2006gq,Sigal:2006bd}. Physically, the two states of the promoter
could reflect changes in the structure of chromatin, the binding of
transcription factors, or stalling of RNA polymerases during
transcription.

\subsubsection*{Typical numbers for constitutive expression}
Some typical numbers for constitutive (unregulated) expression in {\it
  E.\ coli} are
\eqa
d_1= 1/\mbox{hour} &;& d_0 = 1/\mbox{3 minutes} \nonu \\
\la N \ra = 10^3 &;& \la M \ra = 5
\aqe
and so (\ref{etaN}) becomes
\eqa
\eta^2_N &=& 1/1000 + 3/60 \times 1/5 \nonu \\
&=& 0.001 + 0.01
\aqe
The mRNA term determines the overall magnitude of the noise.

\setcounter{equation}{0}
\renewcommand{\theequation}{A\arabic{equation}}

\subsection*{Appendix 1: Dirac delta function}
\begin{figure}[ht]
\begin{center}
\scalebox{0.5}{\includegraphics{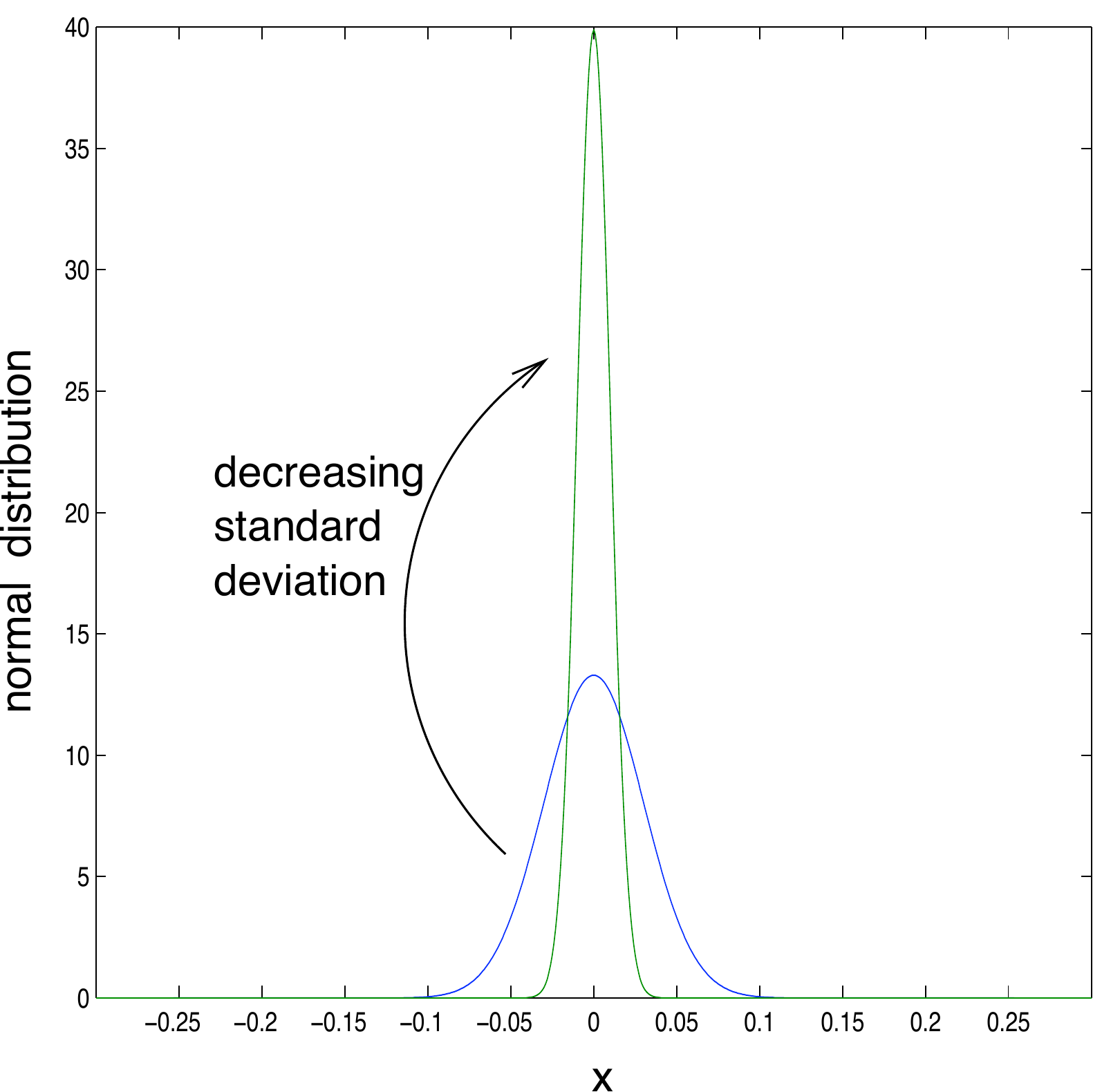}}
\caption{\small The Dirac delta function is the `spike' limit of a normal distribution as its standard deviation tends to zero.} 
\label{delta}
\end{center}
\end{figure}

The Dirac delta function can be considered the limit of a zero mean
normal distribution as its standard deviation tends to zero:
\equ
\delta(x) = \stackrel{\rm lim}{\scriptstyle n \rightarrow \infty} \frac{n}{\sqrt{\pi}} \e^{-n^2 x^2}
\uqe
This limit gives a function whose integral over all $x$ is one, but
that becomes increasingly more and more spiked at zero (Fig.\
\ref{delta}). Ultimately
\equ
\delta(x)= 0 \mbox{ for all $x \ne 0$}
\uqe
and is not strictly defined at $x=0$, but does retain the property
\equ
\int_{-\infty}^{\infty} \delta(x) dx = 1 .
\uqe

These two characteristics imply that the integral of a product of a
delta function and another function, $f(x)$, will only give a non-zero
result at $x=0$. The delta function effectively selects the value
$f(0)$ from the integral:
\equ
\int_{-\infty}^{\infty} f(x) \delta(x) dx = f(0)
\uqe
or more generally
\equ
\int_{-\infty}^{\infty} f(x) \delta(x-y) dx = f(y) .
\uqe

\subsection*{Appendix 2: Sampling from a probability distribution}
\begin{figure}[ht]
\begin{center}
\scalebox{0.5}{\includegraphics{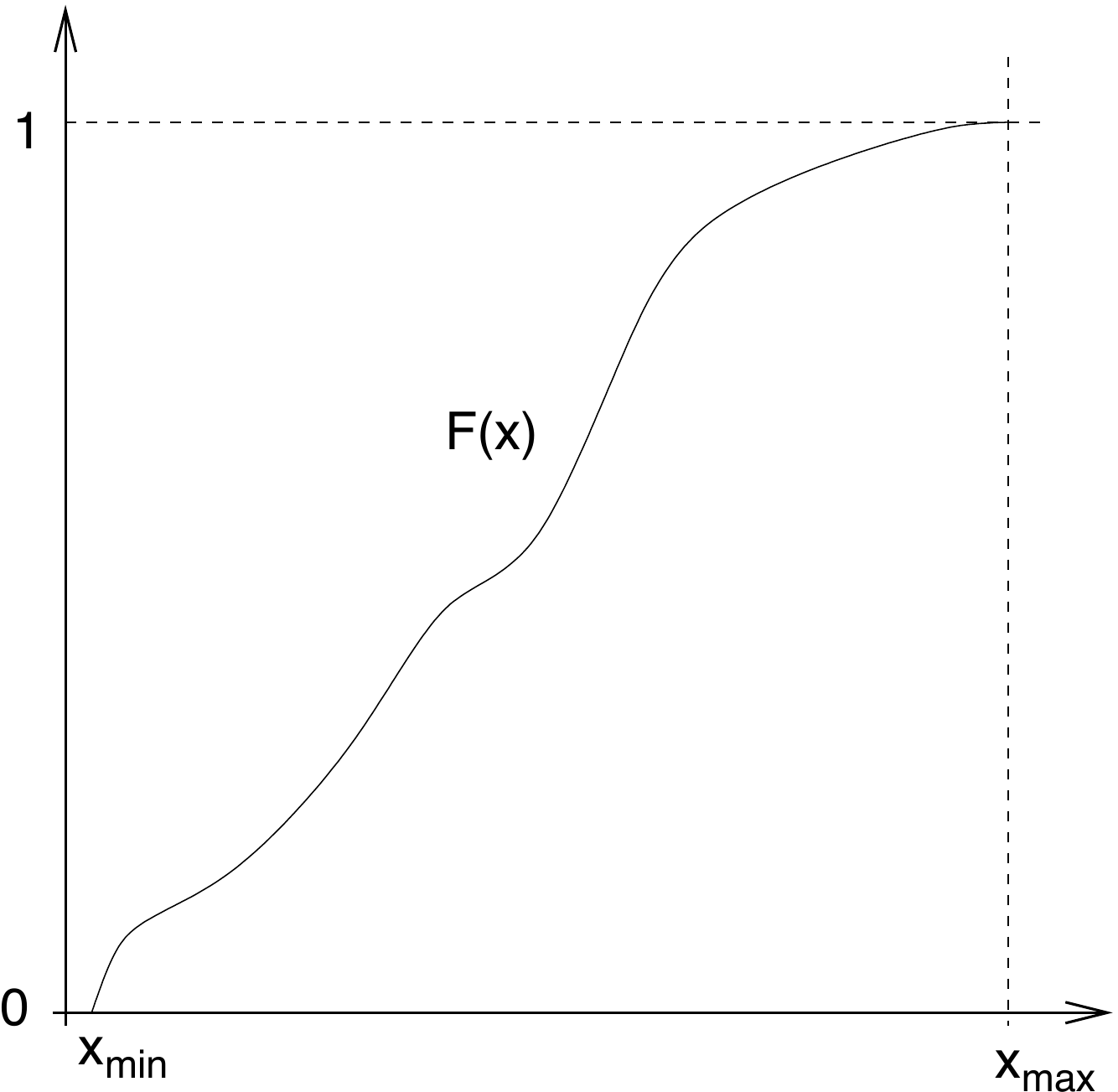}}
\caption{\small A typical plot of cumulative frequency versus $x$.} 
\label{cumu}
\end{center}
\end{figure}

Often we wish to sample from a particular probability distribution,
$P(x)$, say. The cumulative distribution of $P(x)$ is
\equ
F(x) = \int_{x_{\rm min}}^x P(x') dx'
\uqe
and
\eqan{Pdefn}
{\cal P}(x \le x_0) &=& \int_{x_{\rm min}}^{x_0} P(x') dx' \nonu \\
&=& F(x_0)
\naqe
A sketch of the typical behaviour of $F(x)$ is shown in Fig.\
\ref{cumu}. If $x \le x_0$, then $F(x) \le F(x_0)$ because $F(x)$ is a
monotonic increasing function (by definition).

To sample from $P(x)$, first let $y$ be a uniform random number with
$0 \le y \le 1$ (easily obtained on a computer), then
\equn{uni}
{\cal P}(y \le y_0) = \int_0^{y_0} dy' = y_0
\nuqe
for some $0 \le y_0 \le 1$. Define 
\equn{xdef}
x= F^{-1}(y)
\nuqe
where $F(x)$ is the cumulative frequency of $P(x)$. Consequently,
\eqa
{\cal P}(x \le x_0) &=& {\cal P}( F^{-1}(y) \le x_0 ) \nonu \\
&=& {\cal P}(F.F^{-1}(y) \le F(x_0))
\aqe
given that $F(x)$ is monotonic. As $F.F^{-1}(y)= y$, we have
\eqa
{\cal P}(x \le x_0) &=& {\cal P}(y \le F(x_0)) \nonu \\
&=& F(x_0)
\aqe
as $y$ is a sample between 0 and 1 from the uniform distribution: see
(\ref{uni}).  Thus the $x$ of (\ref{xdef}) obeys (\ref{Pdefn}) and so
is a sample from $P(x)$.

If we can calculate the inverse function of the cumulative frequency
of a distribution $P(x)$, then applying this inverse function to a
sample from the uniform distribution gives a sample from $P(x)$.


\end{document}